\documentclass[aps,showpacs]{revtex4}
\usepackage{graphicx}
\def\bc{\begin{center}}
\def\ec{\end{center}}

\def\beq{\begin{equation}}
\def\eeq{\end{equation}}

\begin{document}

\title{Dynamical creation of entangled bosonic states in a double well 
}

\author{K. Ziegler}
\affiliation{Institut f\"ur Physik, Universit\"at Augsburg, D-86135 Augsburg, Germany}
\date{\today}

\begin{abstract}
We study the creation of a bosonic N00N state from the evolution 
of a Fock state in a double well. While noninteracting bosons disappear quickly in the 
Hilbert space, the evolution under the influence of a Bose-Hubbard Hamiltonian is much 
more restricted. This restriction is caused by the fragmentation of 
the spectrum into a high-energy part with doubly degenerate levels and a nondegenerate 
low-energy part. This degeneracy suppresses transitions to states of the high-energy 
part of the spectrum. At a moderate interaction strength this 
effect supports strongly the dynamical formation of a N00N state.
The N00N state is suppressed in an asymmetric double well, where the double degeneracy 
is absent. 

\end{abstract}

\maketitle

\section{Introduction}

Recent experiments on ultracold gases in optical potentials \cite{trotzky08,esteve08,gross10} 
and experiments on photons in microwave cavities \cite{brune08,wang08} have demonstrated that it 
is possible to prepare a Fock state as a pure
state in a finite-dimensional system. After the preparation of the Fock state, the parameters of the system 
can suddenly be changed (performing a ``quench'') such that the Fock state is not an eigenstate of the new 
system Hamiltonian $H$. Then the evolution of the many-body state due to the evolution operator $\exp(-iHt)$ 
will lead to a random walk inside the available Hilbert space. The visited states include other Fock states
as well as superpositions of Fock states. Typical questions in this context are: what is 
the probability for visiting different states and how is this affected by the interaction of the 
particles? A natural quantity for measuring this probability is the 
spectral density function of the Hamiltonian $H$ with respect to the initial Fock state 
\cite{ziegler10a,annibale10,ziegler10b}. 

A classical candidate for modeling the evolution of a Fock state is the Hubbard model \cite{hubbard63,lewenstein07}. 
The corresponding discrete Hamiltonian describes the tunneling of a particle between 
neighboring potential wells and a local particle-particle interaction.
The Hubbard model for bosons (Bose-Hubbard model) was realized as an atomic system in an optical
lattice \cite{greiner02}.
A possible realization of the Bose-Hubbard model by photons in coupled microwave cavities 
was proposed recently by Hartmann et al. \cite{hartmann}. An anharmonicity of the microwave 
cavities plays the role of the photon-photon interaction \cite{ziegler10b}.

The simplest system for discussing the evolution of a Fock state within the Hubbard model 
is a double well, where particles can tunnel between the two wells. For $N$ bosons the
underlying Hilbert space is spanned by the $(N+1)$--dimensional Fock base 
$\{ |0,N\rangle , |1,N-1\rangle ,..., |N,0\rangle\}$, where $l$ bosons are in one well and 
$N-l$ in the other well \cite{milburn97,smerzi97,mahmud03,salgueiro07}. 
The initial state is prepared as a Fock state, where all the bosons are in one of the two wells 
(i.e. $|0,N\rangle$ or $|N,0\rangle$), while the tunneling between the 
wells is turned off. To start the evolution, a ``quench'' is provided by switching on the 
tunneling between the two wells. This is realized by a sudden reduction the potential barrier 
between the wells in an atomic system \cite{trotzky08} or by connecting the two microwave cavities 
with an optical fiber \cite{ji09,hartmann,ziegler10b}.
A similar experiment was performed with two atomic clouds, subject to weak  
interaction and separated by an adjustable potential barrier \cite{albiez05,gati06}. 

On the theoretical side, mean-field descriptions of the Bose-Hubbard model, 
such as a Hartree approximation or the Gross-Pitaevskii equation, 
may work well for clouds with many bosons and weak boson-boson interaction \cite{smerzi97}.
However, they provide a rather poor approximation for
the dynamics of small many-body systems (cf. Ref. \cite{milburn97}). This was also observed
in a recent study by Streltsov et al. who compared the results of a simple Hartree (Gross-Pitaevskii) 
approximation with a sophisticated  multi-orbital Hartree approximation \cite{streltsov09}.
The latter reveals that the bosonic clouds 
are related to superpositions of Fock states in the form of N00N states
\beq
|N00N\rangle=\frac{1}{\sqrt{2}}\left[|0,N\rangle+e^{i\phi N}|N,0\rangle\right]
\ .
\label{noon}
\eeq
In the following we will study the Hubbard dynamics of bosons in a double well in more detail.
In particular, we are interested in the connection of spectral properties and the 
formation of N00N states, based on a Fock state with all the particles in one well as the initial state.
To avoid problems with uncontrolled approximations, we will rely on a full quantum calculation. An exact solution 
is available 
in a Fock-state base, as described previously in Refs. \cite{ziegler10a,ziegler10b}.

The paper is organized as follows: In Sect. \ref{sect:model} the model, based on the Bose-Hubbard Hamiltonian,
is defined and in Sect. \ref{sect:dynamics} the dynamics of an isolated quantum system is explained. 
Then we discuss the dynamics of a noninteracting Bose gas in Sect. \ref{sect:double_well_non} and the dynamics 
of an interacting Bose gas in Sect. \ref{sect:double_well_int}. The latter is divided into a study of
a symmetric double well (Sect. \ref{sect:double_well_symm}) and of an asymmetric double well 
(Sect. \ref{sect:double_well_asymm}). Finally, we summarize the results of our calculation in 
Sect. \ref{sect:results} and discuss them in Sect. \ref{sect:discussion}.

\section{Model}
\label{sect:model}

The many-body Hamiltonian ${\hat H}$ of $N$ bosons with mass $m$ reads 
\beq
{\hat H}=\sum_{j=1}^N 
\left[\frac{{\bf p}_j^2}{2m}+V({\bf r}_j)\right]
+\sum_{j,k=1}^NU({\bf r}_j,{\bf r}_k)
\ ,
\label{mb_ham0}
\eeq
where ${\bf p}_j$ is the momentum of a boson,
$V({\bf r}_j)$ is the one-body potential of the double well and $U$ is 
the two-body interaction potential. For the latter we assume that it decays very quickly 
with the distance $|{\bf r}_j-{\bf r}_k|$ of the particles. This implies that particles
located in different wells do not interact with each other.
Then the many-body Hamiltonian is expressed in Fock-state representation as
\[
\int\cdots\int
\langle N-k,k|{\bf r}_1, {\bf r}_2, ..., {\bf r}_N\rangle
\langle {\bf r}_1, {\bf r}_2, ..., {\bf r}_N|{\hat H}|{\bf r}_1', {\bf r}_2', ..., {\bf r}_N'\rangle
\langle|{\bf r}_1', {\bf r}_2', ..., {\bf r}_N'|N-k',k'\rangle
d^3{\bf r}_1\cdots d^3{\bf r}_N'
\]
\beq
=\langle N-k,k|H|N-k',k'\rangle
\ .
\label{bhh0}
\eeq
For the new Hamiltonian $H$, which acts in the Hilbert space spanned by the Fock base,
we can use the Bose-Hubbard Hamiltonian with local interaction in each well as a reasonable approximation
\beq
H=J(a_1^\dagger a_2+a_2^\dagger a_1)+U_1(a_1^\dagger a_1)^2+U_2(a_2^\dagger a_2)^2
\ ,
\label{bhh}
\eeq
where $a_j^\dagger$ ($a_j$) are creation (annihilation) operators for bosons in the Fock states.
$H$, which describes tunneling between the two wells and the local interaction inside the well with 
interaction strength $U_j$, 
gives us a complete quantum description of the different Fock states and their superpositions. 
In particular, we can use it to study the evolution of a Fock state to a N00N state of Eq. (\ref{noon}).

\subsection{Evolution of isolated systems}
\label{sect:dynamics}

We consider a system which is isolated from the environment.
Furthermore, we assume that the system lives in an ($N+1$)--dimensional Hilbert space. 
With the initial state $|\Psi_0\rangle$  
we can get for the time evolution of the state
\beq
|\Psi_t\rangle=e^{-iHt}|\Psi_0\rangle
\label{evol0}
\eeq
or the evolution of the return probability $|\langle\Psi_0|\Psi_t\rangle|^2$ with the amplitude
\beq
\langle\Psi_0|\Psi_t\rangle =\langle\Psi_0|e^{-iHt}|\Psi_0\rangle 
\ .
\label{return0}
\eeq
In general, the amplitude $\langle\Psi_1|\Psi_t\rangle$
can be expressed via an integral transformation of the resolvent as
\beq
\langle\Psi_1|\Psi_t\rangle =\langle\Psi_1|e^{-iHt}|\Psi_0\rangle
=\int_\Gamma \langle\Psi_1|(z-H)^{-1}|\Psi_0\rangle e^{-izt}dz
\ ,
\label{laplace}
\eeq
where the contour $\Gamma$ encloses all the eigenvalues $E_j$ ($j=0,1,...,N$) of $H$.
With  the corresponding eigenstates $|E_j\rangle$ the spectral representation of the resolvent is a rational function:
\beq
\langle\Psi_1|(z-H)^{-1}|\Psi_0\rangle=\sum_{j=0}^N\frac{\langle\Psi_1|E_j\rangle\langle E_j|\Psi_0\rangle}{z-E_j}
=\frac{P_N(z)}{Q_{N+1}(z)}
\ ,
\label{resolvent2}
\eeq
where $P_N(z)$, $Q_{N+1}(z)$ are polynomials in $z$ of order $N$, $N+1$, respectively, with 
the common denominator
\[
Q_{N+1}(z)=\prod_{j=0}^N(z-E_j)
\ .
\]
These polynomials are readily evaluated by the recursive projection method (RPM) \cite{ziegler10a}. 

The expression in Eq. (\ref{resolvent2}) for $|\Psi_1\rangle=|\Psi_0\rangle$ can be interpreted 
as the bosonic spectral density $\rho_\epsilon(E)$ with respect to the state $|\Psi_0\rangle$:
\beq
\rho_\epsilon(E)=\frac{1}{\pi}Im \langle \Psi_0|(E-i\epsilon-H)^{-1}|\Psi_0\rangle
= \frac{\epsilon}{\pi}\sum_{j=0}^N\frac{|\langle\Psi_0|E_j\rangle|^2}{\epsilon^2+(E-E_j)^2}
\ .
\label{spectrald}
\eeq
The amplitude of the return probability then reads as the Fourier transform of the spectral density
\beq
\langle\Psi_0|\Psi_t\rangle
=\lim_{\epsilon\to 0}\int\rho_\epsilon(E)e^{-iEt}dE
\ .
\label{fourier}
\eeq
Analogously, the overlap $\langle\Psi_1|\Psi_t\rangle$ reads in terms of the resolvent
\beq
\langle\Psi_1|\Psi_t\rangle = \frac{1}{\pi}\lim_{\epsilon\to 0}
\int Im \langle\Psi_1|(E-i\epsilon-H)^{-1}|\Psi_0\rangle e^{-iEt}dE
\label{fourier2}
\eeq
with
\beq
\lim_{\epsilon\to 0} Im \langle\Psi_1|(E-i\epsilon-H)^{-1}|\Psi_0\rangle
=\pi\sum_j\langle\Psi_1|E_j\rangle\langle E_j|\Psi_0\rangle\delta(E-E_j) 
\ ,
\label{off_diagonal1}
\eeq
provided that the matrix elements are symmetric. The latter is the case for the  Hubbard Hamiltonian.

The purpose of the subsequent calculation is to determine the evolution of the Fock
state under the influence of the Bose-Hubbard Hamiltonian of Eq. (\ref{bhh}). In general,
this is expressed in the Fock base as
\beq
|\Psi_t\rangle=\sum_{j=0}^N c_j(t)|N-j,j\rangle
\label{evolution1}
\eeq
with coefficients $c_j(t)=\langle N-j,j|\Psi_t\rangle$. 
For the N00N state we only need to focus on the coefficients $c_0(t)$ and $c_N(t)$.

Comparing the result in Eq. (\ref{evolution1}) with the expressions in Eqs. (\ref{fourier}), (\ref{fourier2}), (\ref{off_diagonal1}),
it turns out that the Fourier transform of $c_0(t)$ and $c_N(t)$ are just the imaginary parts of the matrix elements
of the resolvent
\beq
{\tilde c}_0(E)=\frac{1}{\pi}\lim_{\epsilon\to 0}Im \langle N,0|(E-i\epsilon-H)^{-1}|N,0\rangle 
=\sum_j\langle N,0|E_j\rangle\langle E_j|N,0\rangle\delta(E-E_j) 
\label{fourier_c1}
\eeq
and
\beq
{\tilde c}_N(E)=\frac{1}{\pi}\lim_{\epsilon\to 0}Im \langle 0,N|(E-i\epsilon-H)^{-1}| N,0\rangle 
=\sum_j\langle 0,N|E_j\rangle\langle E_j|N,0\rangle\delta(E-E_j) 
\ .
\label{fourier_c2}
\eeq
These two expressions will be called spectral coefficients, where ${\tilde c}_0(E)$ measures the relative weight
$|\langle N,0|E_j\rangle|^2$. Integration over the energy $E$ gives 1 for this coefficient. 
The coefficient ${\tilde c}_N(E)$ measures the correlation between $|N,0\rangle$ and $|0,N\rangle$
due to the product $\langle 0,N|E_j\rangle\langle E_j|N,0\rangle$. The latter is real for a symmetric Hamiltonian.
Integration over the energy $E$ gives 0 for this coefficient.

\section{Double well: Noninteracting Bose gas}
\label{sect:double_well_non}

The Bose-Hubbard Hamiltonian has two simple limits: The local limit $J=0$ and the noninteracting limit $U_1=U_2=0$.
In the local limit for a symmetric double well with $U_1=U_2$ pairs Fock states $|N-k,k\rangle$, $|k,N-k\rangle$ 
are doubly degenerate eigenstates with 
energy $E_k=U[(N-k)^2+k^2]$. A perturbation by a small tunneling term will break the degeneracy. This effect
is stronger at lower energies because the parabolic spectrum is denser there. This agrees with a numerical 
study \cite{milburn97}. The fact that the states $|N,0\rangle$ and $|0,N\rangle$ are very close in energy may 
support the formation of a N00N state.   

In the absence of particle-particle interaction the Bose-Hubbard Hamiltonian $H_t$
(i.e. the Hamiltonian in Eq. (\ref{bhh}) with $U_1=U_2=0$) describes only tunneling.
A straightforward calculation shows that the eigenstate $|N-k;k\rangle$ of $H_t$ with 
$H_t|N-k;k\rangle=J(N-2k)|N-k;k\rangle$ has an overlap with the Fock states $|N,0\rangle$ and $|0,N\rangle$ as
\beq
\langle N,0|N-k;k\rangle = 2^{-N/2}\sqrt{{N \choose k}} , \ \ \
\langle 0,N|N-k;k\rangle = (-1)^k 2^{-N/2}\sqrt{{N \choose k}}
\ .
\label{binomi1}
\eeq
This implies that the spectral coefficients of Eqs. (\ref{fourier_c1}) and (\ref{fourier_c2})
have a binomial form
\beq
{\tilde c}_0(E)= 2^{-N}\sum_{k=0}^N{N \choose k}\delta(E+J(2k-N))
\label{free_bose3a}
\eeq
\beq
{\tilde c}_N(E)= 2^{-N}\sum_{k=0}^N{N \choose k}(-1)^k\delta(E+J(2k-N))
\ .
\label{free_bose3b}
\eeq
A Fourier transformation reveals a periodic behavior of the evolutionary coefficients as
\beq
c_0(t) 
=\langle N,0|e^{-iHt}|N,0\rangle=\cos^N(Jt) , \ \ \
c_N(t) 
=\langle 0,N|e^{-iHt}|N,0\rangle=(-i)^N\sin^N(Jt)
\ .
\label{non_int_exp}
\eeq
Thus the evolution of the Fock state leads to a N00N state with a 
probability that decays exponentially with $N$. This is a consequence
of the fact that for an increasing $N$ the particles disappear in the $(N+1)$--dimensional
Hilbert space because there is no constraint due to interaction.

\section{Double well: interacting Bose gas}
\label{sect:double_well_int}

The double well with the two Fock states $|N,0\rangle$, $|0,N\rangle$ as possible initial states
can be treated within the RPM. This method is based on a systematic expansion of
the resolvent $\langle \Psi_1|(z-H)^{-1}|\Psi_0\rangle$, starting from the initial base $\{|N,0\rangle,|0,N\rangle\}$.
The method can also be understood as a directed random walk in Hilbert space. This means that in comparison
with the conventional random walk the directed random walk of the RPM visits a subspace ${\cal H}_{2j}$ only once 
and never returns to it. In terms of $N$ bosons, distributed over the double well, the subspace ${\cal H}_{2j}$
is spanned by the base $\{|N-j,j\rangle,|j,N-j\rangle\}$.  A step  from ${\cal H}_{2j}$ to ${\cal H}_{2j+2}$ 
is given by the Hamiltonian $H$ in such a way that ${\cal H}_{2j+2}$ is created by acting $H$ on ${\cal H}_{2j}$ (cf. App. A). 
This step is provided by the tunneling of a single boson. Thus, the directed random walk follows a path with increasing
numbers $j$. The directed random walk is the main advantage of the RPM which allows us to calculate the matrix elements
$\langle \Psi_0|(z-H)^{-1}|\Psi_0\rangle$, $\langle \Psi_1|(z-H)^{-1}|\Psi_0\rangle$ of the resolvent
on a $(N+1)$-dimensional Hilbert space exactly.

\subsection{Symmetric double well}
\label{sect:double_well_symm}

Now we choose $U_1=U_2 \equiv U$ for the Bose-Hubbard Hamiltonian.
Assuming that $N$ is even, all projected spaces ${\cal H}_{2j}$ are two-dimensional
and spanned by $\{|N-j,j\rangle , |j,N-j\rangle\}$ ($j=0,...,N/2$). This leads to a recurrence relation
in the base of the two Fock states 
$
\left(|N,0\rangle ,\ \ \ |0,N\rangle\right)
$
as initial states. The recurrence relation reads (App. A)
\beq
g_{k+1}=\pmatrix{
a_{k+1} & b_{k+1} \cr
b_{k+1} & a_{k+1} \cr
} , \ \ 
g_0=\frac{1}{z-UN^2/2}\pmatrix{
1 & 0 \cr
0 & 1 \cr
} \ \ \ (k=0,1,...,N/2-1)
\label{2d_recursion}
\eeq
with coefficients
\beq
a_{k+1}=\frac{z-{\tilde f}_{k+1}-J^2a_k(N/2+k+1)(N/2-k)}
{\left[z-{\tilde f}_{k+1}-J^2a_k(N/2+k+1)(N/2-k)\right]^2-J^4b_k^2(N/2+k+1)^2(N/2-k)^2}
\label{coeff_a}
\eeq
\beq
b_{k+1}=\frac{J^2b_k(N/2+k+1)(N/2-k)}
{\left[z-{\tilde f}_{k+1}-J^2a_k(N/2+k+1)(N/2-k)\right]^2-J^4b_k^2(N/2+k+1)^2(N/2-k)^2}
\label{coeff_b}
\eeq
and
\[
{\tilde f}_{k+1}=U(N/2+k+1)^2+U(N/2-k-1)^2
\ .
\]
The iteration terminates after $N/2$ steps with
\beq
g_{N/2}=
\pmatrix{
a_{N/2} & b_{N/2} \cr
b_{N/2} & a_{N/2} \cr
}
\ ,
\label{termination}
\eeq
where
\beq
a_{N/2} = \langle N,0|(z-H)^{-1}|N,0\rangle = \langle 0,N|(z-H)^{-1}|0,N\rangle
\ ,
\label{fock1}
\eeq
and 
\beq
b_{N/2} =\langle 0,N|(z-H)^{-1}|N,0\rangle = \langle N,0|(z-H)^{-1}|0,N\rangle
\ .
\label{fock2}
\eeq

There exists an invariance of the recurrence relation under the following simultaneous sign 
changes in Eqs. (\ref{coeff_a}) and (\ref{coeff_b}) 
\beq
z\to -z, \ \ U\to -U, \ \ a_j\to -a_j, \ \ b_j\to -b_j
\ .
\label{transf}
\eeq
This implies that a change from a repulsive to an attractive Hubbard interaction results in a mirror
image with respect to energy of the spectral coefficients 
\beq
{\tilde c}_0(E,U)={\tilde c}_0(-E,-U) ,\ \ \ {\tilde c}_N(E,U)={\tilde c}_N(-E,-U)
\ .
\label{image}
\eeq

\subsection{Asymmetric double well}
\label{sect:double_well_asymm}

In the case $U_1=-U_2\equiv U$ we have one more variable, namely $a_k$, $b_k$ and $c_k$ with
the following recurrence relations (App. A)
\beq
g_{k+1} =\pmatrix{
a_{k+1} & c_{k+1} \cr
c_{k+1} & b_{k+1} \cr
} , \ \ g_0=\frac{1}{z}
\pmatrix{
1 & 0 \cr
0 & 1 \cr
}\ \ \ (k=0,1,...,N/2-1)
\label{2d_recursion1}
\eeq
with matrix elements ($n=N/2$):
\beq
a_{k+1}=\frac{z+4Un(k+1)-J^2(n+k+1)(n-k)b_k}{D_{k+1}}
\label{acoeff_a}
\eeq
\beq
b_{k+1}=\frac{z-4Un(k+1)-J^2(n+k+1)(n-k)a_k}{D_{k+1}}
\label{acoeff_b}
\eeq
\beq
c_{k+1}=-\frac{J^2(n+k+1)(n-k)c_k}{D_{k+1}}
\label{acoeff_c}
\eeq 
and with
\[
D_{k+1}=[z-4Un(k+1)-J^2a_k(n+k+1)(n-k)][z+4Un(k+1)-J^2b_k(n+k+1)(n-k)]
\]
\[
-J^4c_k^2(n+k+1)^2(n-k)^2
\ .
\]
The final result of the iteration is
\beq
g_{N/2}=
\pmatrix{
a_{N/2} & b_{N/2} \cr
b_{N/2} & c_{N/2} \cr
}
\ ,
\label{termination2}
\eeq
with
\beq
a_{N/2}=\langle N,0|(z-H)^{-1}|N,0\rangle , \ \ \
b_{N/2}=\langle 0,N|(z-H)^{-1}|0,N\rangle 
\ ,
\label{fock3a}
\eeq
\beq
c_{N/2}=\langle N,0|(z-H)^{-1}|0,N\rangle=\langle 0,N|(z-H)^{-1}|N,0\rangle
\ .
\label{fock3b}
\eeq

\section{results}
\label{sect:results}

\begin{figure}
\begin{center}
\includegraphics[width=10cm,height=7cm]{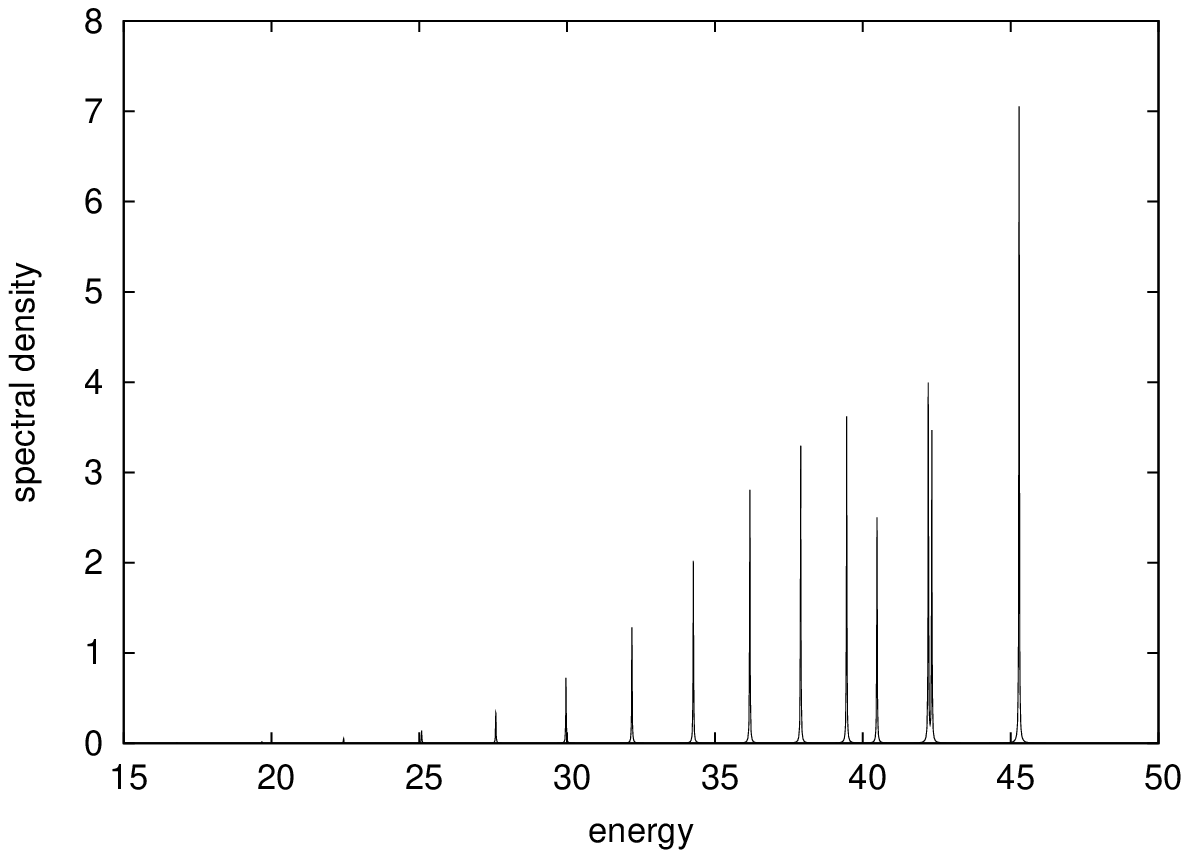}
\includegraphics[width=10cm,height=7cm]{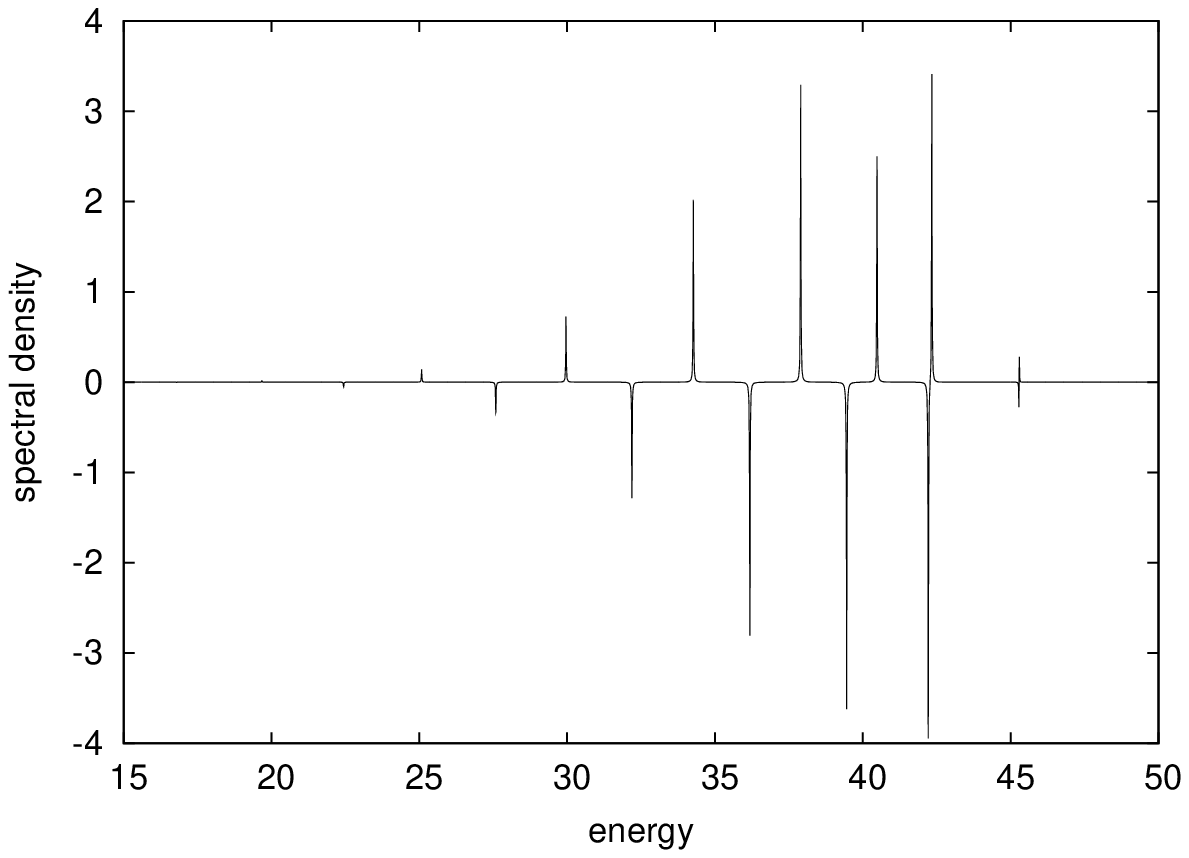}
\caption{
Spectral coefficients for 20 bosons with $U/J\approx 0.1$ and $\epsilon=0.01$.
The energy of the initial Fock state is ${\bar E}=40$. The almost degenerate states appear above $42$.
}
\label{fig:3}
\end{center}
\end{figure}

\begin{figure}
\begin{center}
\includegraphics[width=10cm,height=7cm]{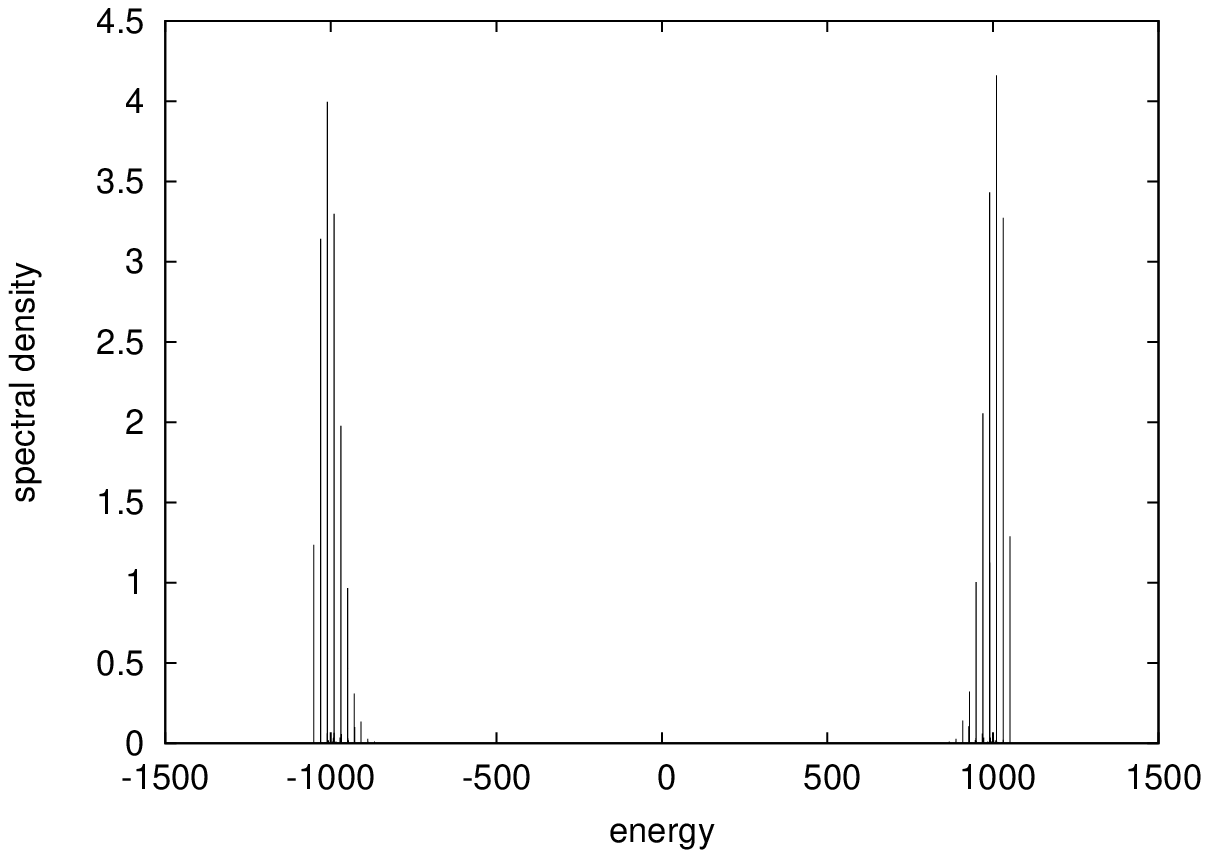}
\includegraphics[width=10cm,height=7cm]{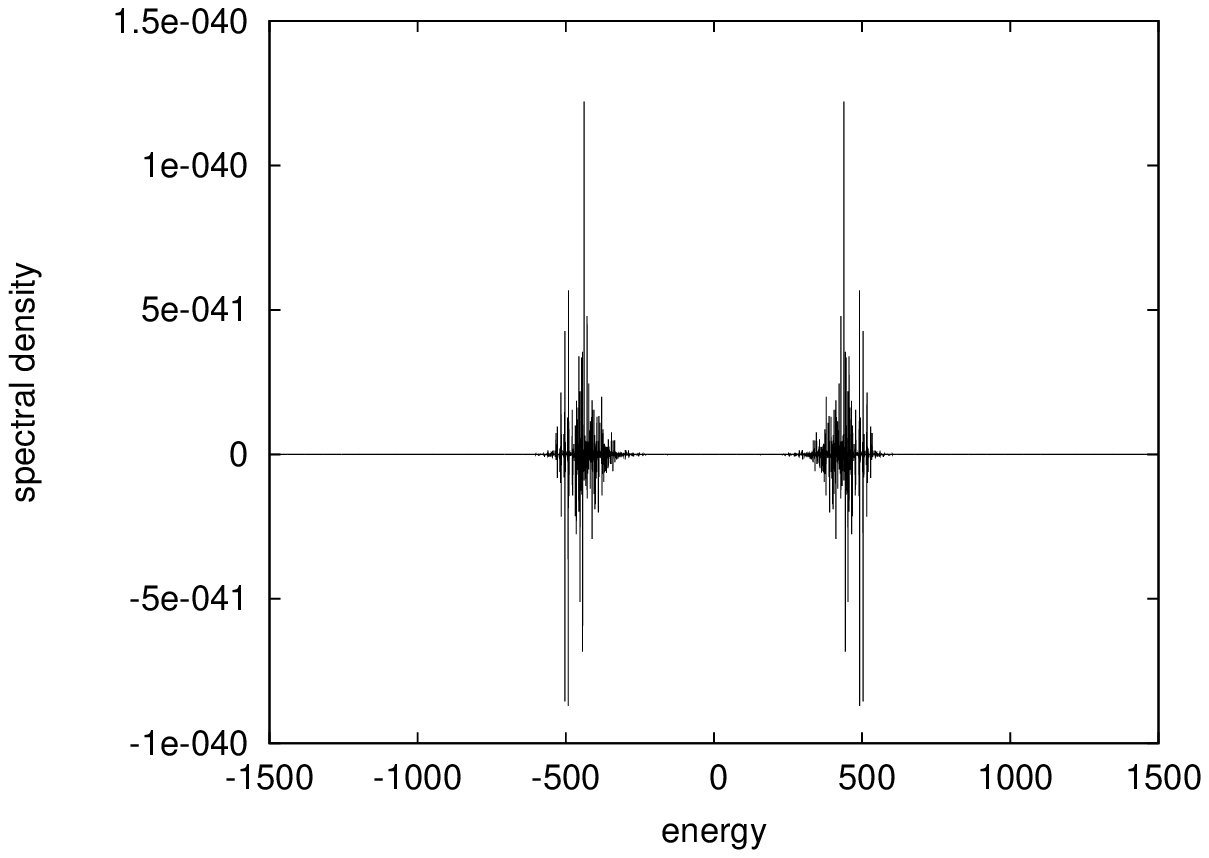}
\caption{
Spectral coefficients 
for 100 bosons in an asymmetric double well with $U/J\approx \pm 0.023$ and 
$\epsilon=0.01$. 
The energy of the Fock states $|0,N\rangle$ and $|N,0\rangle$ is ${\bar E}=\pm 1000$, respectively.}
\label{fig:6}
\end{center}
\end{figure}

The iteration of Eqs. (\ref{coeff_a}), (\ref{coeff_b}) for a symmetric double well and the iteration
of Eqs. (\ref{acoeff_a})-(\ref{acoeff_c}) for an asymmetric double well
gives us, according to Eqs.  (\ref{fock1}), (\ref{fock2}) and  (\ref{fock3a}), (\ref{fock3b}),
the following matrix elements of the resolvent
\[
\langle N,0|(z-H)^{-1}|N,0\rangle ,\ \   
\langle 0,N|(z-H)^{-1}|N,0\rangle = \langle N,0|(z-H)^{-1}|0,N\rangle
\ .
\]
These are rational functions of $z$, as shown in Eq. (\ref{laplace}).
For N bosons these are lengthy expressions with $N+1$ poles. Therefore, it is convenient
to present the results as plots with respect to energy.
Examples of the spectral coefficients ${\tilde c}_0(E)$ and ${\tilde c}_N(E)$
are shown for a symmetric double well with 100 bosons in Fig. \ref{fig:1} and with 20 bosons 
in Fig. \ref{fig:3}, and for an asymmetric double well with 100 bosons in Fig. \ref{fig:6}. 
A larger number of bosons shows a richer spectral structure. The diagonal coefficient 
${\tilde c}_0(E)$ in the case of 100 bosons is remarkably different from the 
off-diagonal coefficient ${\tilde c}_N(E)$ because the latter does not have 
spectral weight from eigenstates whose energy $E_j$ is larger than the energy of the initial Fock
state ${\bar E}=UN^2$. The reason for this feature is the double degeneracy of the eigenvalues
mentioned in Sect. \ref{sect:double_well_non}: 
The signs of the product $\langle 0,N|E_j\rangle\langle E_j|N,0\rangle$ for adjacent eigenvalues 
are opposite to each other. Since the eigenvalues get closer pairwise as we increase their energy, the 
contribution of the two levels cancel each other for each pair inside the sum of Eq. (\ref{fourier_c2}). 
This interaction effect is also visible for 20 bosons (Fig. \ref{fig:3}), although the cancellation 
is incomplete then due to a larger level distance.
This can be considered as an effect of spectral fragmentation, where the spectrum has a nondegenerate
low-energy and a degenerate high-energy part, caused by the competition of
tunneling and interaction. 

\begin{figure}
\begin{center}
\includegraphics[width=10cm,height=7cm]{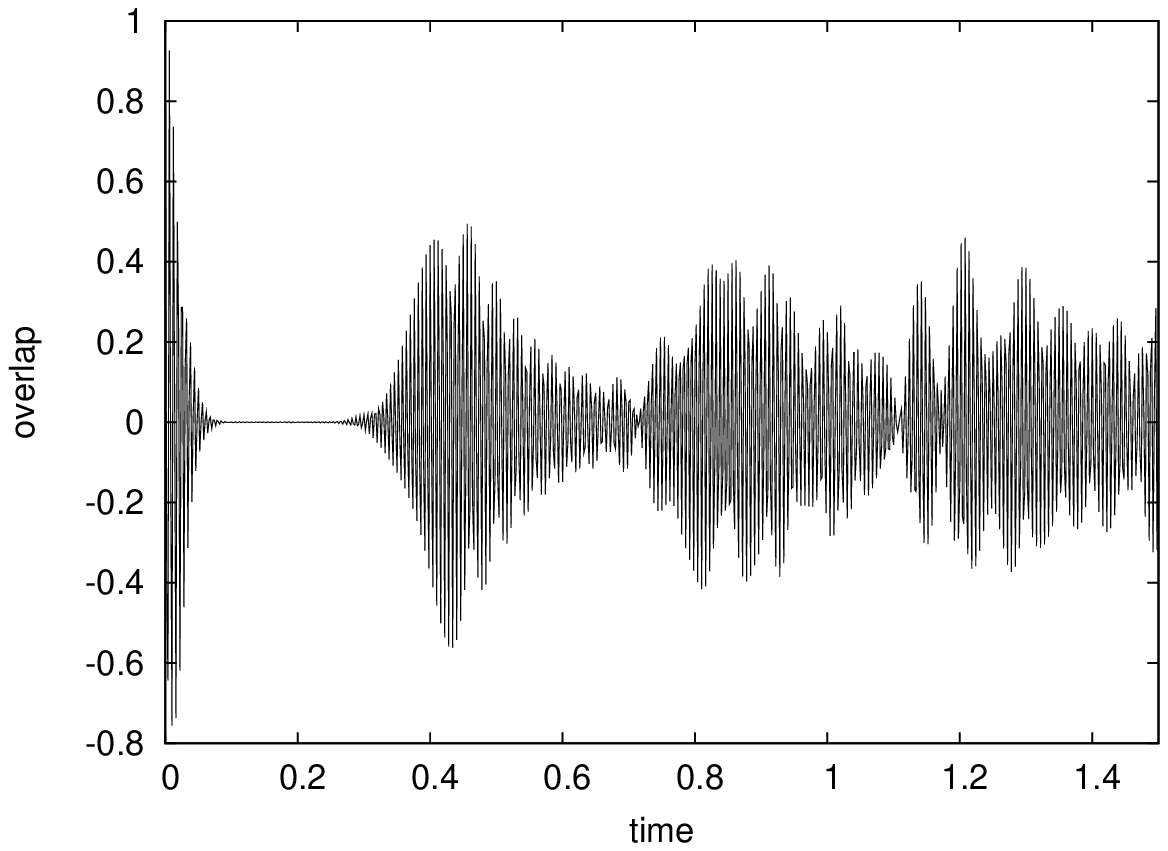}
\includegraphics[width=10cm,height=7cm]{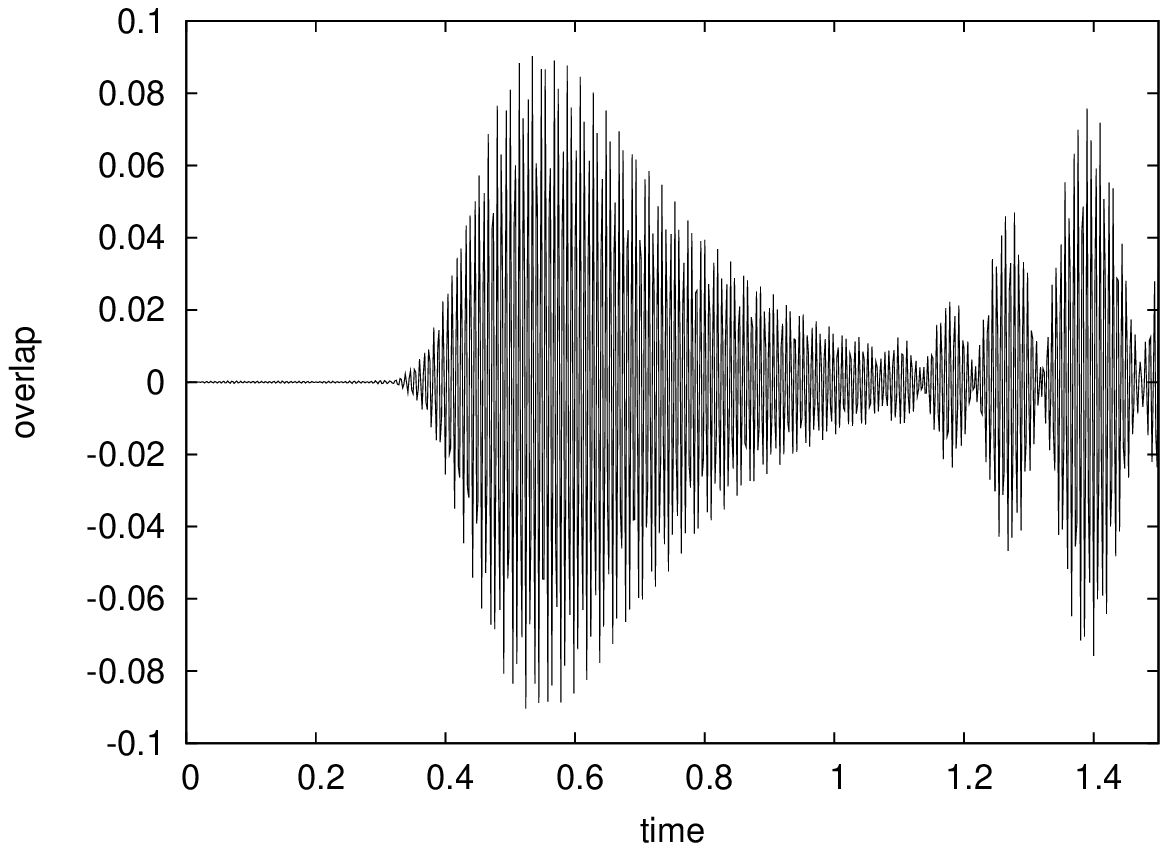}
\caption{
Evolution of the real part of the evolutionary coefficients $c_0(t)=\langle N,0|\Psi_t\rangle$ (upper panel), 
$c_N(t)=\langle 0,N|\Psi_t\rangle$ (lower panel) for 100 bosons with $U/J\approx 0.023$.
The time scale is given in inverse units of the interaction strength $0.1\hbar/U$ and the 
energy of the initial state is ${\bar E}=1000$.
}
\label{fig:2}
\end{center}
\end{figure}

\begin{figure}
\begin{center}
\includegraphics[width=10cm,height=7cm]{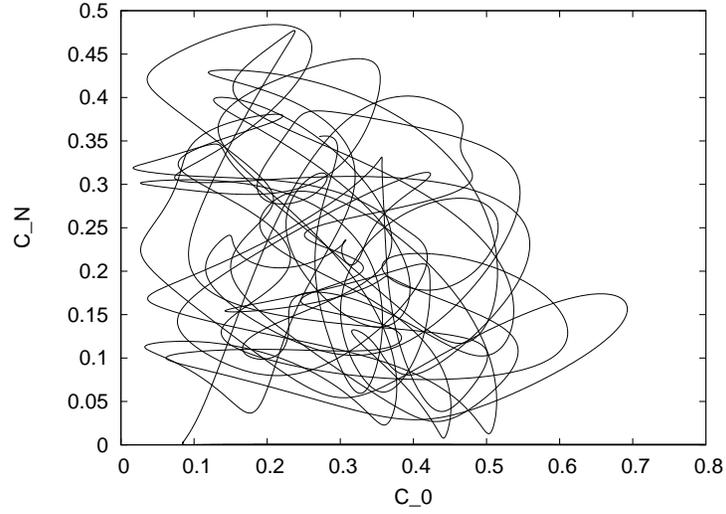}
\caption{
Evolution of $C_0=|c_0(t)|$, $C_N=|c_N(t)|$ for 20 bosons with $U/J\approx 0.1$
over a time period of $0.4 \hbar/U$.
The trajectory $(|c_0(t)|,|c_N(t)|)$ starts at $(1,0)$.
}
\label{fig:4}
\end{center}
\end{figure}

\begin{figure}
\begin{center}
\includegraphics[width=12cm,height=7cm]{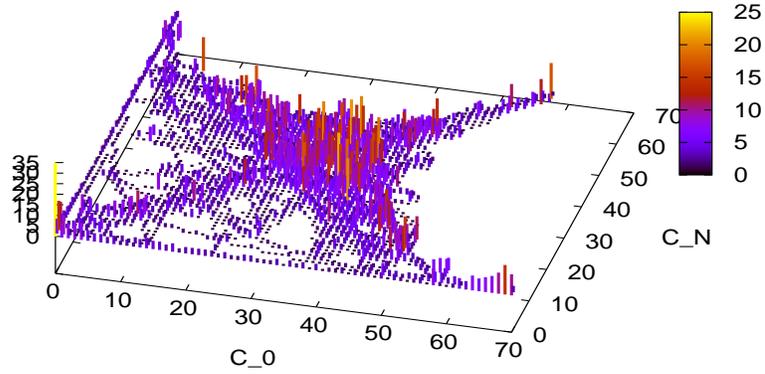}
\caption{
Distribution $P(|c_0(t)|,|c_N(t)|)$ of $|c_0(t)|$, $c_N(t)|$
over a time period of $0.4 \hbar/U$ for 20 bosons with $U/J\approx 0.026$.
The axes are scaled by a factor 100 and the vertical axis is in arbitrary units.
This plot indicates a strong correlation between the two spectral coefficients, supporting
the formation of a N00N state.
}
\label{fig:5}
\end{center}
\end{figure}

The contribution of the two Fock states $|0,N\rangle$, $|N,0\rangle$ to the
evolution in Eq. (\ref{evolution1}) is given by the coefficients 
$c_0(t)=\langle N,0|\Psi_t\rangle$, $c_N(t)=\langle 0,N|\Psi_t\rangle$. 
In Fig. \ref{fig:2} the real parts of these coefficients are plotted for 100
bosons. Their evolution indicates a collapse and revival behavior. The plot of $|c_0(t)|$, $|c_N(t)|$ as a two-dimensional vector
for 20 bosons in Fig. \ref{fig:4} shows a complex dynamical behavior that cannot be described by a simple equation of motion. 
This observation suggests a statistical description with a probability $P(|c_0(t)|,|c_N(t)|)$ which measures how often certain
values of $|c_0(t)|$, $|c_N(t)|$ are visited during the evolution in a period of time.  The result for 20 bosons is plotted in
Fig. \ref{fig:5} for $U=0.1$, $J=\sqrt{15}$. It indicates that there is a strong correlation between the coefficients,
where the most favored values are $|c_0(t)|\approx |c_N(t)|\approx 0.35$.   

For an asymmetric double well with interaction strength $\pm U$ the spectrum is different because of the absence
of double degeneracy of the eigenvalues (cf. Fig. \ref{fig:6}). 
There are two ``bands'', one around $E=UN^2$, the other around $E=-UN^2$, where the widths of the bands
is characterized by the tunneling rate $J$.  Moreover, the off-diagonal part ${\tilde c}_N(E)$ appears closer 
to zero energy and its values are very small. This indicates that the off-diagonal part has overlaps with energy 
levels which are different from those of the diagonal part ${\tilde c}_0(E)$ . For the evolution only the latter 
contribute substantially, preventing the system to create a N00N state.

\section{Discussion and Conclusions}
\label{sect:discussion}

In order to understand the evolution of an isolated many-body bosonic system, we start with noninteracting
bosons (i.e. $U_1=U_2=0$) of Sect. \ref{sect:double_well_non}.  The spectral properties are characterized by (i) equidistant
energy levels with distance $J$ and (ii) a binomial weight distribution of the energy levels. The evolution of a Fock state
is characterized by a periodic behavior with a single frequency $\omega=J/2\pi$ as a direct consequence of the
equidistant energy levels. The amplitudes for visiting the initial Fock state $|N,0\rangle$ or the complimentary
Fock state $|0,N\rangle$ vary with $\cos^N (Jt)$ or $(-i)^N\sin^N (Jt)$, respectively. This implies for a large number $N$
of bosons that (i) these states are visited only for a very short period of time and (ii) the two Fock states  are visited
at different times. Thus the formation of a N00N state is very unlikely for noninteracting bosons.  

\begin{figure}
\begin{center}
\includegraphics[width=10cm,height=7cm]{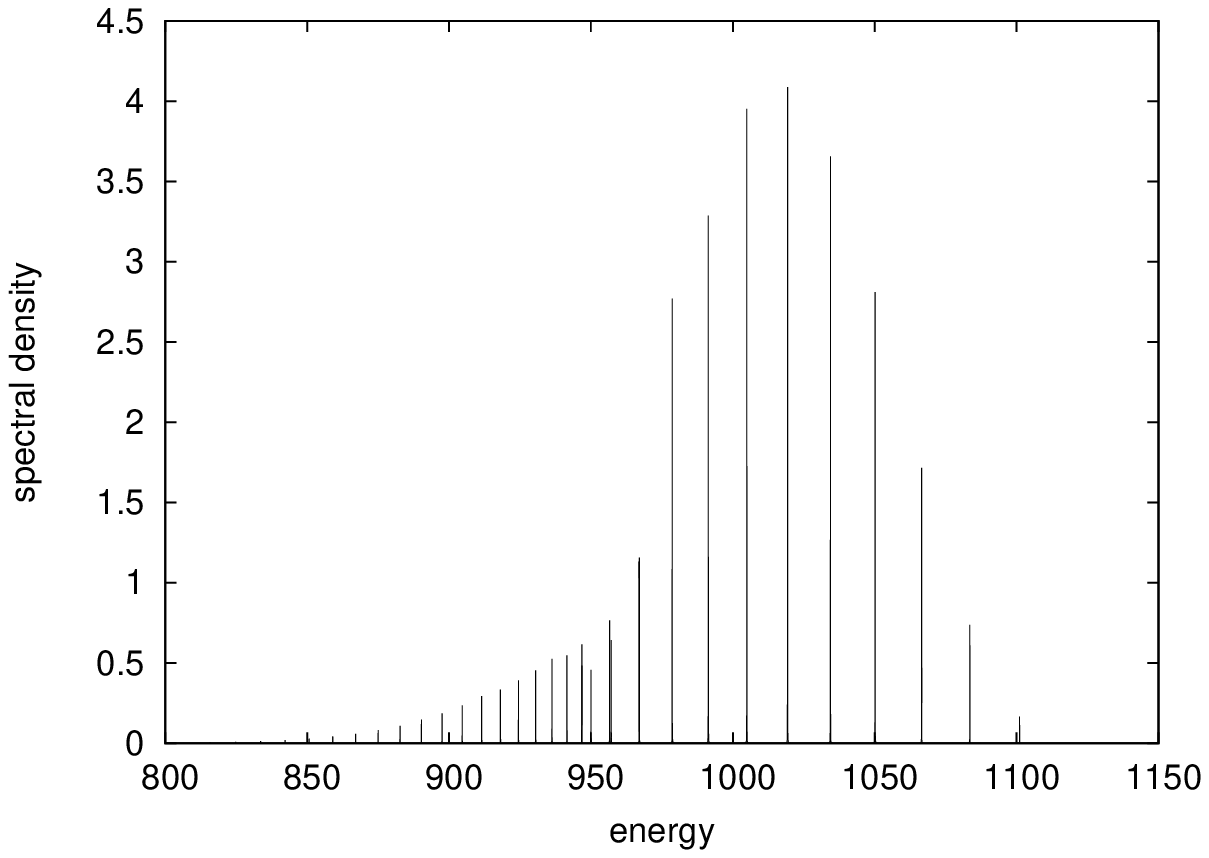}
\includegraphics[width=10cm,height=7cm]{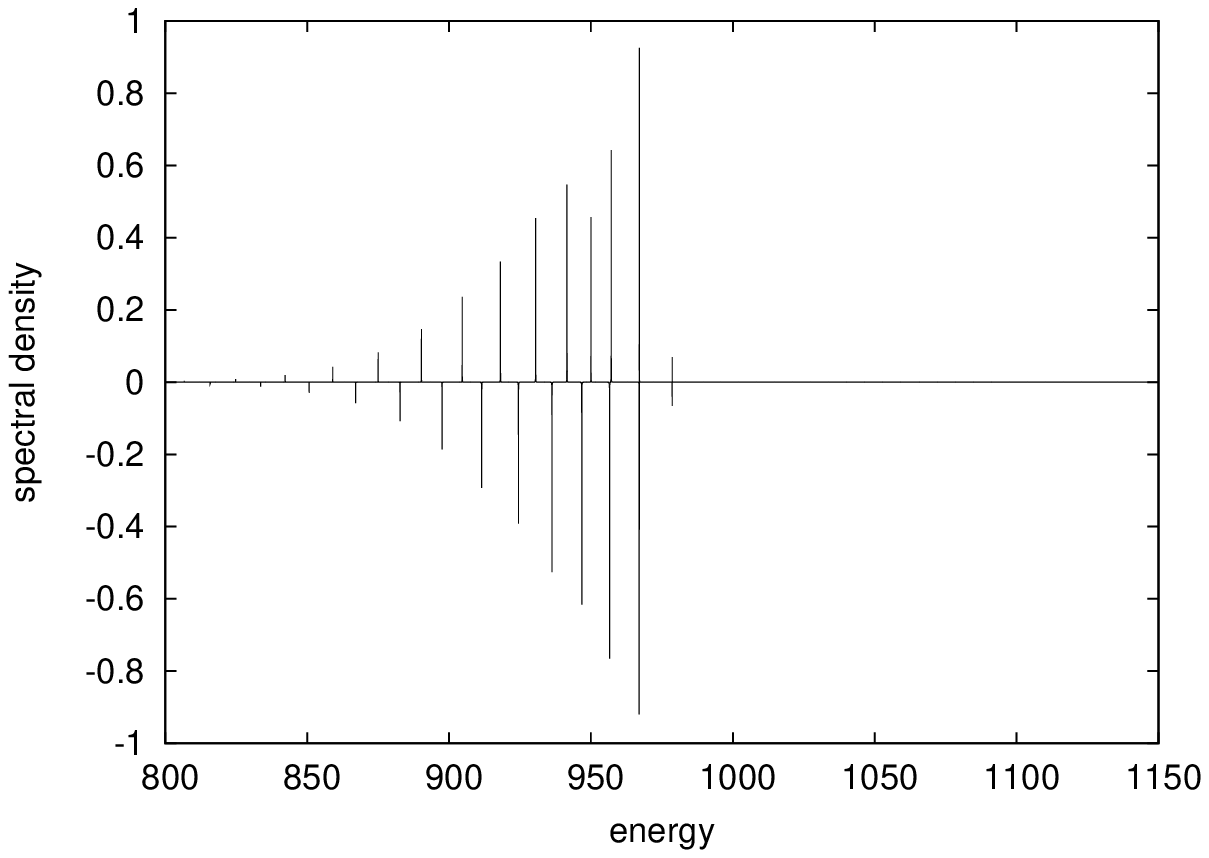}
\caption{
Spectral coefficients of Eqs. (\ref{fourier_c1}) and (\ref{fourier_c2})
for 100 bosons with $U/J\approx 0.023$ and 
$\epsilon=0.01$. The energy of the initial Fock state is ${\bar E}=1000$.
The spectral fragmentation appears around $E\approx 970$, where the levels 
are nondegenerate at lower energies but almost degenerate for higher energies.
This is a consequence of the competition between tunneling and interaction,
which the latter wins at higher energies.
}
\label{fig:1}
\end{center}
\end{figure}

A simple qualitative picture for the general evolution of the Fock state is the random walk in Hilbert space. In case of
noninteracting bosons the particles can walk independently of each other which enables them to explore the
entire Hilbert space spanned by the Fock states without restriction. A simultaneous overlap of $|\Psi_t\rangle$ with both 
Fock states $|N,0\rangle$ and $|0,N\rangle$ is very unlikely then. Once we have turned on the boson-boson
interaction the particles experience a mutual influence which restricts their individual random walks. This is related
to the fact that the systems stays much longer in the energetically (almost) degenerate Fock states $|N,0\rangle$ 
and $|0,N\rangle$ than in the noninteracting
case (cf. Fig. \ref{fig:4}) and, what is even more important here, they can have a simultaneous overlap with both
Fock states, such that they create a N00N state. In terms of the spectral properties the interaction modifies (i) the 
energy levels, which are not equally spaced, and (ii) the weight distribution of the levels, which are not binomial any longer
(cf. Fig. \ref{fig:1}-\ref{fig:6}). This, of course, affects also the evolution of the Fock state which is more complex now, 
since many different frequencies are involved.  A particular feature is the spectral fragmentation (cf. Fig. \ref{fig:1}),
where only a part of the spectrum contributes to the off-diagonal coefficient ${\tilde c}_N(E)$. This is a kind of Hilbert-space
localization, where transitions to the high-energy part of the Hilbert space are completely suppressed, similar to the self-trapping
found in the Hartree approximation of the Bose-Hubbard model \cite{milburn97}. It should be noticed, however, that
spectral fragmentation appears at a much weaker interaction strength than the self-trapping effect. 
For $U\approx J$, which is the threshold for self-trapping \cite{milburn97}, there is only one eigenvalue with significant weight 
$|\langle N,0|E_j\rangle|^2$ \cite{ziegler10a}.
Thus it is not clear whether or not the two effects are directly are connected. 

For the asymmetric double well the situation is different due to the existence of two ``bands'' and the absence
of the double degeneracy. The main consequence is the absence of a support for the formation of N00N states 
because the off-diagonal coefficient $c_N(t)$ is strongly suppressed. From this observation we can conclude that 
the evolutionary entanglement is much more favorable in the symmetric double well. 
This is in agreement with the results of the multi-orbital Hartree calculation of Ref. \cite{streltsov09}.

In conclusion, we have studied the evolution of a bosonic Fock state $|N,0\rangle$ in a double well and found 
that a local particle-particle interaction supports the formation of a N00N state, provided that the
interaction is not too strong. This is accompanied by a fragmentation of the spectrum. The latter is 
characterized by the fact that only eigenstates with energies less than the energy of the initial Fock 
state can be reached in the evolution. This interaction effect causes a Hilbert-space localization and 
prevents the evolution of the Fock state to disappear in the depth of the Hilbert space. This is the main
reason for a favorable creation of a N00N state. 
The appearance of a N00N state is suppressed though for strong interaction because then the restriction
of the Hilbert space is too severe and does not allow to reach the complementary Fock state $|0,N\rangle$.

\begin{acknowledgements}

I am grateful to A. Streltsov for discussing his work on the N00N state.

\end{acknowledgements}


\appendix

\section{Recursive projection method}

Given is a sequence of projectors $P_j$ ($n\ge j\ge0$), defined by the recurrence relation
\[
P_{2k+1}=P_{2k-1}-P_{2k}\ \ \ (n\ge k\ge0)
\]
with initial conditions $P_{-1}={\bf 1}$, $P_0$ and by the Hamiltonian $H$ through the properties
\beq
P_{2k}HP_{2k+1}=P_{2k}HP_{2k+2},\ \ \
P_{2k+1}HP_{2k}=P_{2k+2}HP_{2k}
\ .
\label{recurse}
\eeq
The projection of the resolvent $(z-H)^{-1}$ defines
\beq
g_k=P_{2(n-k)} \left(z-H\right)_{2(n-k)-1}^{-1} P_{2(n-k)} \ \ \ (0\le k <n)
\ ,
\label{proj_g}
\eeq
where $\left(...\right)_{2(n-k)}^{-1}$ is the inverse on the $P_{2(n-k)}$--projected Hilbert space.
Then $g_k$ satisfies the recurrence relation
\beq
g_k=\left(z-h_k\right)_{2(n-k)}^{-1}
\label{projected4a}
\eeq  
with
\beq
h_k=\cases{
P_{2n}HP_{2n} & $k=0$ \cr
P_{2(n-k)}HP_{2(n-k)}+P_{2(n-k)}Hg_{k-1}HP_{2(n-k)} & $1\le k\le n$ \cr
}
\ .
\label{effham1}
\eeq
Of interest is here only the case $k=n$, where we have from Eq. (\ref{proj_g})
\[
g_n 
=P_0(z-H)^{-1}P_0
\ .
\]

For the specific case of the double well we choose $n=N/2$ and the projectors
\[
P_0=|N,0\rangle\langle N,0|+|0,N\rangle\langle 0,N|, \ \ P_2=|N-1,1\rangle\langle N-1,1|+|1,N-1\rangle\langle 1,N-1|,...,
\]
\[
P_N=|N/2,N/2\rangle\langle N/2,N/2|
\ .
\]
With the Hubbard Hamiltonian of Eq. (\ref{bhh})
the diagonal terms of the effective Hamiltonian in Eq. (\ref{effham1}) read
\[
P_{2(n-k)}HP_{2(n-k)}=P_{N-2k}HP_{N-2k}
\]
\[
=\left[U_1(N/2+k)^2+U_2(N/2-k)^2\right]|N/2+k,N/2-k\rangle\langle N/2+k,N/2-k|
\]
\[
+\left[U_1(N/2-k)^2+U_2(N/2+k)^2\right]|N/2-k,N/2+k\rangle\langle N/2-k,N/2+k|
\ .
\]
The off-diagonal terms of the effective Hamiltonian in Eq. (\ref{effham1}) read 
\[
P_{2(n-k)}HP_{2(n-k+1)}=P_{2N-2k}HP_{N-2k+2}
\]
\[
=-J\sqrt{N/2+k}\sqrt{N/2-k+1}
\]
\[
\left(|N/2+k,N/2-k\rangle\langle N/2+k-1,N/2-k+1|+|N/2-k,N/2+k\rangle\langle N/2-k+1,N/2+k-1|\right)
\ .
\]
This leads for $U_1=U_2$ to Eqs. (\ref{coeff_a}), (\ref{coeff_b}) and for 
$U_1=-U_2$ to Eqs. (\ref{acoeff_a}), (\ref{acoeff_b}) and (\ref{acoeff_c}).


\end{document}